\documentclass[aps,prev,twocolumn,preprintnumbers,floatfix,nofootinbib]{revtex4-1}
\pdfoutput=1
\usepackage[english]{babel}
\usepackage{graphicx,color}
\usepackage{bm}
\usepackage{times}
\usepackage{slashed}
\usepackage{multirow}
\usepackage[usenames,dvipsnames,svgnames,table]{xcolor}
\usepackage{slashed}
\usepackage{amsmath}
\usepackage{amsthm}
\usepackage{subfigure}
\usepackage{hyperref}
\hypersetup{
     linktocpage=true,
     setpagesize=true,
     urlcolor=blue,
     citecolor=blue,
     linkcolor=blue, 
     menucolor=cyan,
     colorlinks=true, 
     filecolor=magenta,
     citebordercolor=blue,
     pdftitle={Vector-Like Fermions and Inert Scalar Triplet},
     pdfsubject={Latex},
     pdfauthor={Alvaro, Yoxara, Farinal, Jose},
    pdfkeywords={BSM, BSM, g-2},
    unicode=true
}

\usepackage[capitalize]{cleveref}
\newcommand{\amu}{\Delta a_{\mu}}
\newcommand{\gmu}{g_{\mu}-2}
\newcommand{\be}{\begin{equation}}
\newcommand{\ee}{\end{equation}}
\newcommand{\bea}{\begin{eqnarray}}
\newcommand{\eea}{\end{eqnarray}}

\usepackage{xfrac}

\begin{document}

\title{Vector-Like Fermions and Inert Scalar Solutions to the Muon g-2 Anomaly \\
and collider probes at the HL-LHC and FCC-hh}

\author{\'A. S. de Jesus$^{a,b}$}
\email{alvarosdj@ufrn.edu.br}

\author{F. S. Queiroz$^{a,b,c}$}
\email{farinaldo.queiroz@ufrn.br}

\author{J. W. F. Valle$^{d}$}
\email{valle@ific.uv.es}

\author{Y. Villamizar$^{e}$}
\email{y.villamizar@ufabc.edu.br}

\affiliation{$^a$ International Institute of Physics, Universidade Federal do Rio Grande do Norte,
Campus Universit\'ario, Lagoa Nova, Natal-RN 59078-970, Brazil}
\affiliation{$^b$ Departamento de F\'isica, Universidade Federal do Rio Grande do Norte, 59078-970, Natal, RN, Brasil}
\affiliation{$^c$ Millennium Institute for Subatomic Physics at High-Energy Frontier (SAPHIR), Fernandez Concha 700, Santiago, Chile}
\affiliation{$^d$ AHEP Group, Institut de F\'{i}sica Corpuscular --  CSIC/Universitat de Val\`{e}ncia, Parc Cient\'ific de Paterna.\\ C/ Catedr\'atico Jos\'e Beltr\'an, 2 E-46980 Paterna (Valencia) - Spain}
\affiliation{$^e$ Centro de Ci\^encias Naturais e Humanas, Universidade Federal do ABC, 09210-580, Santo Andr\'e, S\~ao Paulo, Brasil}

\begin{abstract}
 We examine simple models with an inert scalar and vector-like leptons that can explain the recent $\gmu$ measurement reported at FNAL. 
 Prompted by this exciting result, we explore the viability of a simple interpretation and determine the required parameters. We also embed these models within a 
 3-3-1 gauge extension of the Standard Model (SM), showing that the $\gmu$ anomaly can be accommodated in agreement with current data. We also show how our theory can be tested at high-energy colliders such as HL-LHC and FCC-hh.
\end{abstract}


\maketitle

\section{Introduction}
\label{intro}


A new measurement of the muon magnetic moment was reported by the Muon $\gmu$ Experiment at Fermilab. In addition to the increase of statistics using data collected in 2019 (Run-2) and 2020 (Run-3), the collaboration reduced the systematics by a factor of two. The experiment uses 3.1 GeV polarized muons produced at the Fermilab Muon Campus, which are then injected into a storage ring with a radius of 7.1 m. The experiments surpassed the BNL experiment in many ways \cite{Carey:1999dd,Muong-2:2000vdh,Muong-2:2001kxu,Muong-2:2002wip,Muong-2:2004fok}. After much experimental effort, the Muon $\gmu$ Experiment measured the muon's magnetic moment with unprecedented precision. We remind the reader that from the Dirac equation, we find the muon's magnetic moment to be $\vec{m} = g_{\mu} e / (2 m_\mu) \vec{S}$, where the gyromagnetic ratio is given by $g=2$ \cite{Dirac:1928hu}. The quantum corrections to this $g$-factor can be expressed through the muon anomalous magnetic moment, typically described in terms of $a_\mu=\sfrac{(g_{\mu}-2)}{2}$. The old BNL result \cite{Bennett:2002jb,Bennett:2006fi} prompted a lot of attention in the community, because the discrepancy between theoretical predictions and experimental results would possibly hint for new physics \cite{Czarnecki:2001pv,Davier:2004gb}. There are uncertainties in the significance of the value of $\gmu$ due to theoretical issues stemming from high-order SM hadronic contributions, particularly the Hadronic Vacuum Polarization (HVP) and Light-by-Light (HLbL) corrections \cite{Aoyama:2020ynm,Jegerlehner:2017gek,Crivellin:2020zul,Borsanyi:2020mff,Blum:2019ugy,CMD-3:2023alj}. A recently conducted study \cite{Borsanyi:2020mff} has significantly reduced the uncertainty around the Leading-Order Hadronic Vacuum Polarization (LO-HVP) theoretical contribution to the muon anomalous magnetic moment ($\gmu$) from lattice QCD calculations, showing a considerable discrepancy between the lattice and the data-driven approaches. This study combined with other known Standard Model contributions has reduced the discrepancy in the muon g-2 theory to around 2$\sigma$. There is also some tension regarding the data-driven approach. A new result from the CMD-3 collaboration has shown incompatible results compared with past measurements of the $e^+e^- \to \pi^+ \pi^-$ channel, showing that it has a larger cross-section than expected. This process represents the largest contribution to the R-ratio, thus it brings the theoretical result closer to the experimental limit on $a_\mu$, reducing the anomaly in the data-driven front \cite{CMD-3:2023alj}. However, both results from CMD-3 and BMW collaboration are still under scrutiny, with more measurements and calculations needed to either confirm or confront these results. In Table \ref{tableI} we report the discrepancy and its significance to highlight that this $\gmu$ anomaly is far from being resolved.

\begin{table}[h!]
  \centering
  \caption{Summary of the $\gmu$ anomaly over the years.}
  \vspace{0.1cm}
  \rowcolors{2}{gray!10}{white} 
  \begin{tabular}{lccc} 
   \toprule 
       \multicolumn{4}{c}{ \vspace{-0.4cm}  }\\
       Current value to $\gmu$ & standard deviation &  year & Reference\\
       $\Delta a_\mu  =  (261 \pm 78)\times 10^{-11}$  &  $3.3\sigma$ & 2009 & \cite{Prades:2009tw,Tanabashi:2018oca}\\
       $\Delta a_\mu  =   (325 \pm 80)\times 10^{-11}$ & $4.05\sigma$ & 2012 & \cite{Benayoun:2012wc}\\
       $\Delta a_\mu  =  (287 \pm 80)\times 10^{-11}$ & $3.6\sigma$ & 2013  &\cite{Blum:2013xva}\\
       $\Delta a_\mu =  (377 \pm 75)\times 10^{-11}$ & $5.02\sigma$ & 2015 &\cite{Benayoun:2015gxa}\\
       $\Delta a_\mu =  (313 \pm 77)\times 10^{-11}$ & $4.1\sigma$ & 2017 &\cite{Jegerlehner:2017lbd}\\
       $\Delta a_\mu = (270 \pm 36)\times 10^{-11}$ & $3.7\sigma$ & 2018 &\cite{Keshavarzi:2018mgv}\\ 
        $\Delta  a_\mu = (250 \pm 48) \times 10^{-11}$& $5.0 \sigma$& 
        2023 & \cite{Muong-2:2023cdq} \\ 
  \end{tabular}
\label{tableI}
\end{table}

The current value given by the Particle Data Group (PDG) is $\Delta a_\mu=a_\mu^{\exp }-a_\mu^{\mathrm{SM}}=251(41)(43) \times 10^{-11}$ \cite{Workman:2022ynf}, but the recent Muon $\gmu$ experiment reports $\Delta a_\mu=(250 \pm 48)\times 10^{-11}$  \cite{Muong-2:2023cdq}.

Prompted by the recent Muon $\gmu$ result hinting at an exciting $5.0\sigma$ discrepancy between theory and experiment, we discuss how to accommodate this anomaly within a simple setup. Several simplified models have been put forth to account for $\gmu$ \cite{Padley:2015uma,Yamaguchi:2016oqz,Yin:2016shg,Endo:2019bcj,  Endo:2020mqz,Endo:2019mxw,Cox:2018vsv,Konar:2017oah, Un:2016hji,An:2015uwa, Chakraborty:2015bsk,Jana:2020pxx}, see \cite{Lindner:2016bgg} for a recent review.


In this work, considering existing restrictions, we are interested in explaining the $\gmu$ anomaly with vector-like fermions and an inert scalar. Several authors have already suggested vector-like fermions in this context \cite{Dermisek:2013gta,Poh:2017tfo,Barman:2018jhz,Borah:2023dhk}. Here we will be particularly interested in vector-like fermion interaction terms involving the muon. We will also embed our simplified scheme into a non-Abelian 3-3-1 gauge extension of the SM to show how the $\gmu$ anomaly can be addressed comprehensively. These models have a long history~\cite{Singer:1980sw,Valle:1983dk,Pisano:1991ee,Frampton:1992wt}, one of their main motivations being to explain the number of particle families through the anomaly cancellation consistency requirement.

Currently, the search for new particles requires the utilization of simplified models. These models serve as a crucial tool to broaden the scope of our search and help us explore a more comprehensive range of theoretical possibilities. They prove to be particularly useful when investigating physics BSM, such as the anomaly \(g_\mu - 2\). By utilizing these simplified models, we can effectively explain our findings while aligning with the experimental data. Moreover, these simplified models serve as a foundation upon which we can build more complex theories, such as the 3-3-1 models.

Our work is structured as follows: in {\it \cref{secII}}, we describe the simplified models featuring inert scalars and vector-like fermions, presenting regions of parameter space where these models can be used to address the anomaly; in {\it \cref{secIII}}, we introduce a non-Abelian gauge extension of the SM, outlining the region of the parameter that can accommodate the $\gmu$ anomaly in agreement with the data from current and future colliders; in {\it \cref{secIV}} we draw our conclusions. 

\section{\label{secII} Simplified Models}

\subsection{Inert scalar}

In order to illustrate how simplified models can potentially address the $\gmu$ anomaly we first introduce an inert scalar doublet ($\phi$) under $SU(2)_L$ with hypercharge $Y_\phi=1$ via the following Lagrangian,
\begin{equation}
    \mathcal{L} =\lambda_{ab} \bar{L}_{aL} \phi e_{bR} + h.c.
    \label{Eqinert1}
\end{equation} where $L_{aL}$ is the SM lepton doublet, $\lambda_{ab}$ is the Yukawa coupling. With this Lagrangian, there are two contributions to $\Delta a_\mu$ coming from the neutral and charged components of the inert scalar doublet. However, the main contribution to $\gmu$ comes from the neutral component, and is given by \cite{Lindner:2016bgg},
%
%
\begin{equation}
\Delta a_\mu=\frac{\lambda^{\prime 2}}{8 \pi^2} \int_0^1 \mathrm{~d} x \sum_b \frac{\left(g^{s}_{2b}\right)^2 P^{+}(x)+\left(g^{p}_{2b}\right)^2 P^{-}(x)}{(1-x)\left(1-x  \lambda^{\prime 2}\right)+x \epsilon_{b}^2 \lambda^{\prime 2}}
\label{eqg2}
\end{equation}
where $P^{ \pm}(x)=x^2\left(1-x \pm \epsilon_b\right)$, $\epsilon_b=\cfrac{m_b}{m_\mu}$, $\lambda^\prime=\cfrac{m_\mu}{m_{\phi}}$, $g^{s}_{2b}$ and $g^{p}_{2b}$ being the scalar and pseudoscalar couplings of this new scalar with the muon and other particles in the same vertex, represented by the subscript $b$.

This could explain the $\gmu$ anomaly for $0.0076 \, \mbox{GeV}^{-1} < y/m_{\phi} < 0.0102 \, \mbox{GeV}^{-1}$. However, if one embeds this scalar doublet within a complete multi-Higgs model, there are a number of constraints that make this solution rather contrived \cite{Wang:2014sda,Hektor:2015zba,Cherchiglia:2017uwv,Wang:2018hnw,Iguro:2019sly}. 

\begin{figure}
    \centering
    \includegraphics[width=\columnwidth]{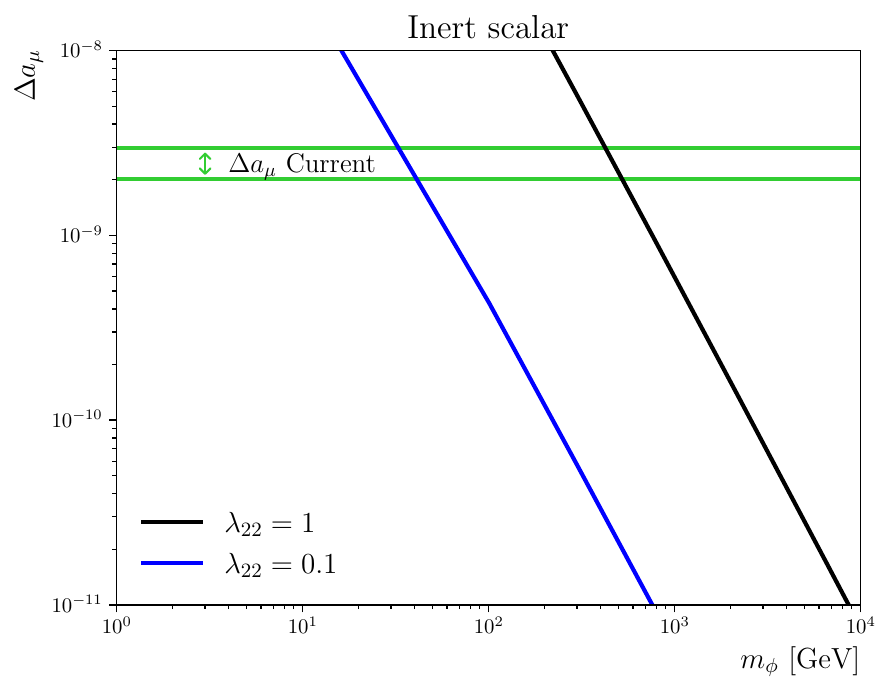}
    \caption{Contribution to $\Delta a_\mu$ from an inert scalar 
    whose mass is a free parameter and whose coupling with the muon is 
    $\lambda_{22} = 1$ (black) and $\lambda_{22} = 0.1$ (blue).}
    \label{inertscalar}
\end{figure}

The inert scalar doublet contribution to $\Delta a_\mu$ is calculated by \cref{eqg2} and shown in \cref{inertscalar}. In this figure, we exhibit the region of $\{\Delta a_\mu, m_\phi\}$ parameter space that provides an adequate contribution to the muon anomalous moment according to the recent data from the $\gmu$ experiment at Fermilab. The latter is illustrated by the band between the horizontal lines in green. We display two different choices for the coupling between the inert scalar and the muon, given by $\lambda_{22} = 1$ and $\lambda_{22} = 0.1$, represented by the solid black and blue lines, respectively. With this setup, both configurations are capable of addressing the anomaly for masses of $m_\phi = 400 - 550$ GeV ($\lambda_{22} = 1$) and $m_\phi = 30 - 40$ GeV ($\lambda_{22} = 0.1$).

\subsection{Vector-like Fermions}

\paragraph{Vector-like fermion plus a scalar singlet}
Vector-like fermions ($E$) represent an interesting class of models to explain the $g-2$ result \cite{Queiroz:2014zfa}. Anomaly cancellation requires vector-like fermions with masses arising from terms of the form $m_E \bar{E_L} E_R$. If $E$ is a singlet under $SU(2)_L$, we can write down,
\begin{equation}
\mathcal{L}= \lambda_{ab} \bar{E}_{aR} \sigma \mu_R + h.c
\label{eq:3}
\end{equation}
where $\sigma$ is now a scalar singlet under $SU(2)_L$.\\

\paragraph{Vector-like fermion plus a scalar doublet}

Notice that we can also generate a positive contribution to $\gmu$ through an exotic charged fermion $(E)$ and inert doublet $(\phi)$ as follows,
\begin{equation}
\mathcal{L}= \lambda_{ab} \bar{L}_{aL} \phi E_{bR} + h.c.
\label{eq:4}
\end{equation}
\paragraph{Vector-like fermion doublet plus an inert scalar doublet}

Finally, we note that we could also add a fermion doublet under $SU(2)_L$ , ($\psi$) plus an inert scalar doublet $\sigma$, and write down,
\begin{equation}
\mathcal{L}= \lambda_{ab} \bar{\psi}_{aL} \sigma \mu_{bR} + h.c.
\label{eq:5}
\end{equation}

\begin{figure}[t]
    \centering
    \includegraphics[width=\columnwidth]{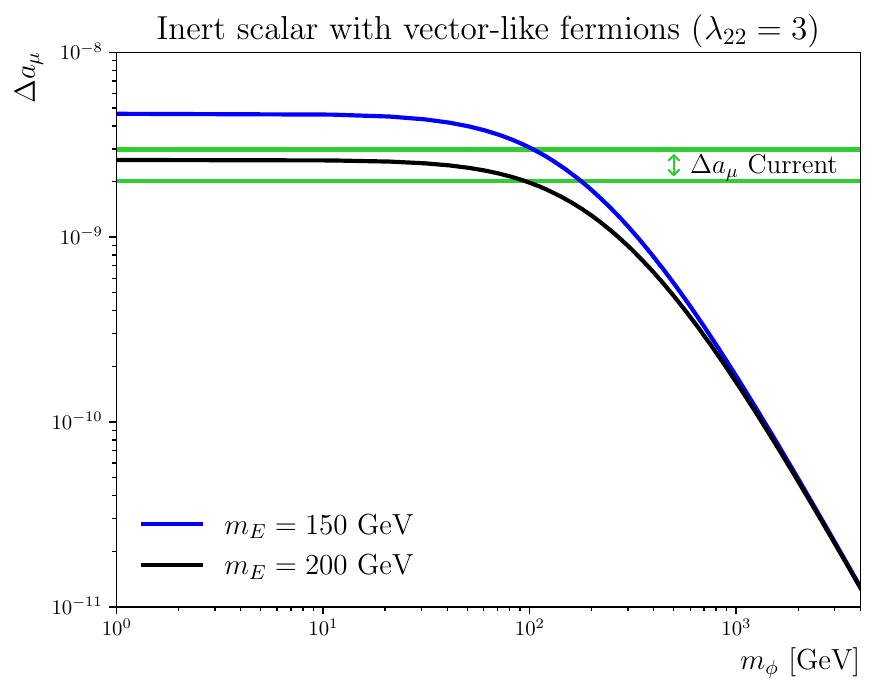}
    \includegraphics[width=\columnwidth]{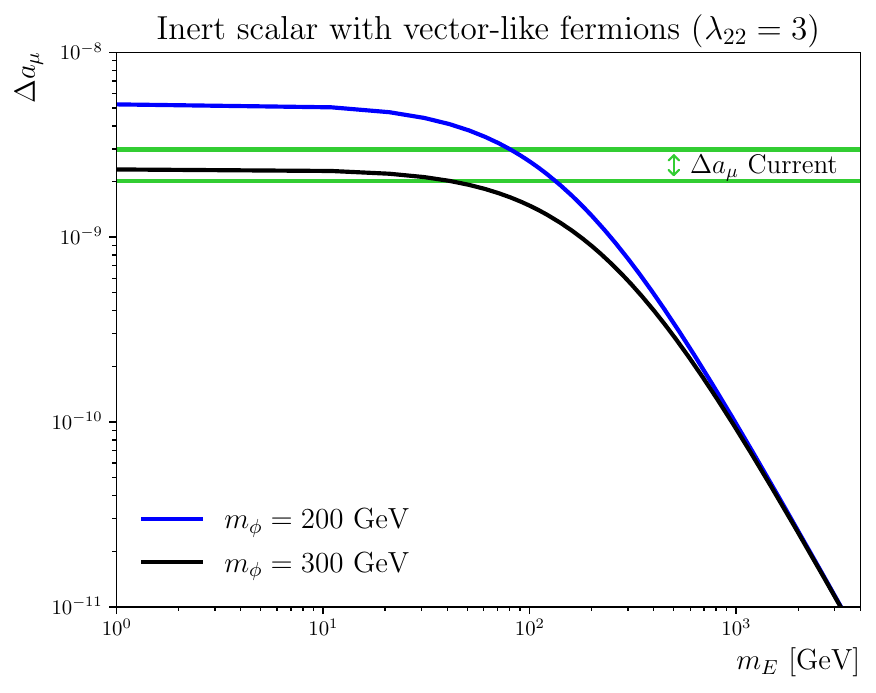}
    \caption{Contributions to $\Delta a_\mu$ from inert scalar and vector-like fermions. The top (lower) panel represents the calculated contribution of a fixed value for the vector-like fermion (scalar) mass. In contrast, the mass of the scalar (vector-like fermion) is a free parameter. The horizontal band between the green lines is the current value.}
    \label{vectorlike}
\end{figure}

\begin{figure}[t]
\centering
\includegraphics[width=1.0\columnwidth]{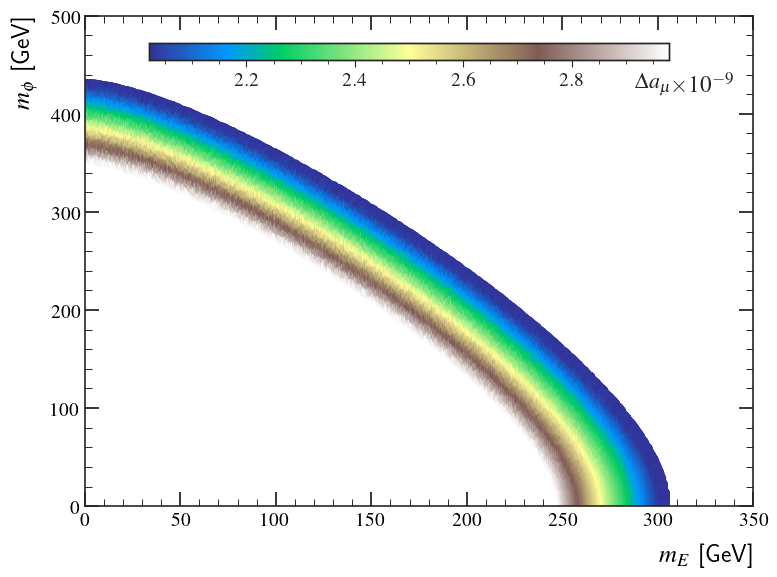}
        \caption{Parameter space exploration in the $m_{\phi} \times m_{E}$ plane, assuming $\lambda_{22} =4$ ($g^{s(p)}_{22} = 2$). The color bar integrated within the plot corresponds to the values of the contribution to $\Delta a_\mu$ inside the allowed region derived with the new $\gmu$ result.}
    \label{mphixme}
\end{figure}

All of these simplified models offer solutions to the $\gmu$ anomaly. We stress that these exotic charged leptons are required to be vector-like in order to ensure gauge anomaly cancellation. The contributions to $g-2$ are determined by Eq.\eqref{eqg2}, where we introduce another free parameter, namely, the mass of the vector-like fermion $m_E$, which impacts the calculation by changing $\epsilon_b = \epsilon_E = m_E/m_\mu$, while in the inert scalar case $\epsilon_b = \epsilon_\mu = 1$. The results from this model are presented in \cref{vectorlike,mphixme}, where the $\gmu$ experiment bounds are represented in the same way as in \cref{inertscalar}, and here we consider three cases: fixed $m_E$ and free $m_\phi$, the opposite configuration, and both $m_E$ and $m_\phi$ as free parameters.

The first two cases are shown in \cref{vectorlike}, with the top (lower) panel representing the first (second) case, both with $\lambda_{22} = 3$. Both cases provide viable solutions to the anomaly, with the top panel representing the contributions to $\Delta a_\mu$ as a function of $m_\phi$ for fixed values of $m_E = 150$ GeV and $m_E = 200$ GeV, while the lower panel shows the contribution to $\Delta a_\mu$ versus $m_E$ assuming $m_\phi = 200$ GeV and $m_\phi = 300$ GeV.

The third case is illustrated in \cref{mphixme}. In this plot we display the region of the $\{m_E, m_\phi\}$ plane leading to an adequate solution to the $\gmu$ anomaly. We find that the contribution to $\Delta a_\mu$ arising from an inert scalar with vector-like leptons can indeed be in agreement with the new $\gmu$ experiment. In this plot, we assume the muon coupling with the scalar and vector-like lepton to be $g_{2b}^{s(p)} = 2$, implying that vector-like fermion masses up to $300$ GeV and inert scalar masses up to $450$ GeV are enough to fit the anomaly. This holds for both the inert scalar and the vector-like lepton model assumed in this work. The horizontal color bar represents the specific value of $\Delta a_\mu$ in units of $10^{-9}$ achieved with each configuration of $m_{\phi}$ and $m_E$, limiting the parameter space to the region inside the new experimental result of $\Delta a_\mu = (250 \pm 48) \times 10^{-11}$. An interesting feature of this plot is the fact that higher values of $m_{\phi}$ imply lower values of $m_E$ and vice-versa. This happens due to the behavior of this channel's contribution to $\Delta  a_\mu$, which depends on the ratio between the masses of the new particles and the muon mass. 

\begin{table}[h!]
  \centering
  \caption{Field content and transformation properties
  }
  \vspace{0.5cm}
  \rowcolors{2}{gray!10}{white}
  \begin{tabular}{lcccc}
   \toprule
   \rowcolor{blue!10!}  
    \multicolumn{5}{c}{Simplified models} \\
    Model & Field & SU(3)$_C$ & SU(2)$_L$ & U(1)$_Y$ \\
    Inert Scalar & $\phi$ & 1 & 2 & 1 \\  
    \multirow{4}{*}{ }Vector-like Fermion & E & 1 & 1 & -2 \\
     & $\phi$ & 1 & 2 & 1 \\
     & $\psi$ & 1 & 2 & -1 \\
     & $\sigma$ & 1 & 2 &  1\\
     \rowcolor{blue!10!}  
    \multicolumn{5}{c}{3-3-1 model} \\
    Model  & Field  & SU(3)$_C$ & SU(3)$_L$ & U(1)$_X$ \\
    3-3-1  & $f$ & 1 & 3 & -1/3 \\
     & $N$ & 1 & 1 &  0\\
    &$\phi$& 1 & 3 &  2/3 \\
  \end{tabular}
  \label{fields}
\end{table}

In our work, we utilize two local symmetry groups, namely $S U(3)_C \times S U(2)_L \times U(1)_Y$ and $S U(3)_C \times S U(3)_L \times U(1)_X$. The new fields and their transformation properties under these symmetry groups are listed in \cref{fields}. Furthermore, in the upcoming \cref{secIII}, we will introduce the 3-3-3 model.

\begin{figure*}
    \centering
    \subfigure[\label{fig1:gmu}]{
    \includegraphics[scale=0.63]{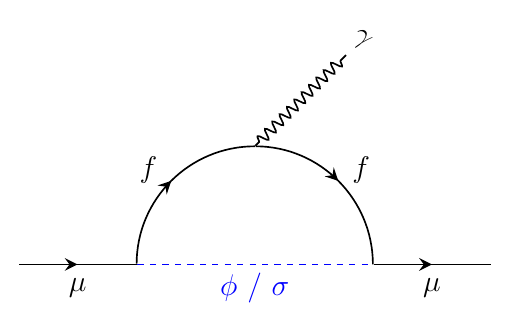}
   }
   \subfigure[\label{fig3:gmu} ]{
    \includegraphics[scale=0.63]{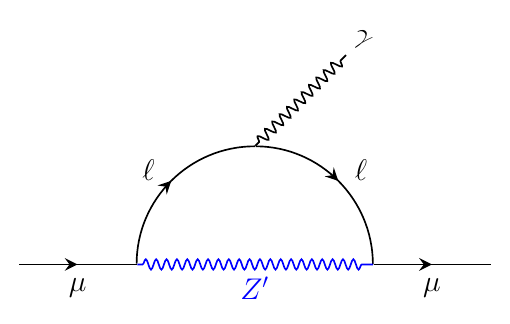}
     }
    \subfigure[\label{fig4:gmu} ]{
    \includegraphics[scale=0.63]{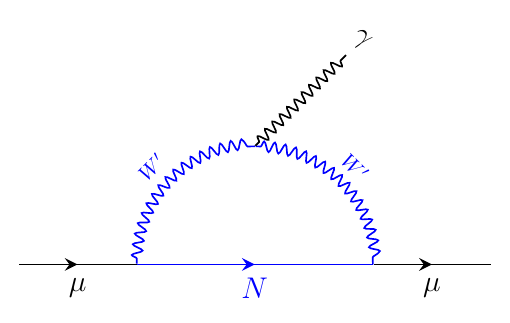}
     }
    \caption{Feynman diagrams contributing to the muon anomalous magnetic 
(loop) in the simplified and 3-3-1 models discussed in this work.}
    \label{fig:gmu}
\end{figure*}

On the other hand, in \cref{fig:gmu} we displayed the Feynman diagrams representations of contributions to the muon's anomalous magnetic moment analyzed in this work. \cref{fig1:gmu} represent the $\gmu$ contributions involving virtual fermions ($f$ represent $\ell = e, \, \mu, \, \tau$ or $E$) and a scalar particle (\(\phi\) or \(\sigma\)) that arise in the simplified and the 3-3-1 models (for instance, see \cref{Eqinert1,eq:3,eq:4,eq:5}). Both scalars are in the simplified models as detailed in the \cref{fields}, whereas, in the 3-3-1 models, we only have the representation of the inert scalar as $\phi$. \cref{fig3:gmu,fig4:gmu} represent the main Feynman diagrams that contribute to the muon anomalous magnetic moment (loop) in the 3-3-1 models with a virtual $Z^{\prime}$ and $W^{\prime}$ boson, and fermions $\ell$ and $N$, respectively (see \cref{eqNC,eqCC}).

\section{\label{secIII}3-3-1 Models}

The so-called 3-3-1 models constitute a broad a interesting class of gauge extensions of the SM based on the $SU(3)_C \times SU(3)_L \times U(1)_X$ symmetry, which has had by now quite a long history \cite{Singer:1980sw,Valle:1983dk,Pisano:1991ee,Frampton:1992wt}. This new gauge symmetry could hold just around the corner, at the TeV scale, while providing a plausible answer to the number of fermion families in the Standard Model as a result of anomaly cancellation. This class of electroweak extensions has a broad range of other possible motivations in Particle Physics. To cite a few, these models have been extensively studied in the context of flavor physics \cite{Cabarcas:2012uf,Santos:2017jbv,Barreto:2017xix,Wei:2017ago,Hue:2017lak}, meson oscillations \cite{CarcamoHernandez:2022fvl,Cogollo:2012ek,Machado:2013jca,Boucenna:2015raa, Kovalenko:2023nhv}, 
charged lepton flavor violation \cite{Boucenna:2015zwa},
collider physics \cite{Meirose:2011cs,Coutinho:2013lta,Nepomuceno:2016jyr,Queiroz:2016gif,Nepomuceno:2019eaz}, neutrino mass generation \cite{Cogollo:2010jw,Cogollo:2008zc,Okada:2015bxa,Vien:2018otl,carcamoHernandez:2018iel,Nguyen:2018rlb,Pires:2018kaj,CarcamoHernandez:2019iwh,CarcamoHernandez:2019vih,CarcamoHernandez:2020pnh,CarcamoHernandez:2020ehn}, the strong CP problem \cite{Dias:2020kbj}, unification  \cite{Boucenna:2014dia,Reig:2016tuk,Hati:2017aez,Deppisch:2016jzl,Dong:2017zxo,Hernandez:2021zje} as well as cosmology. The latter includes the implementation of a viable dark matter scenario \cite{Fregolente:2002nx,Hoang:2003vj,deS.Pires:2007gi,Mizukoshi:2010ky,Profumo:2013sca,Dong:2013ioa,Dong:2013wca,Cogollo:2014jia,Dong:2014wsa,Dong:2014esa,Alves:2016fqe,Carvajal:2017gjj,Montero:2017yvy,Huong:2019vej,Kang:2019sab,Leite:2019grf,CarcamoHernandez:2020ehn,Leite:2020wjl,Leite:2020bnb,Alvarado:2021fbw}, as well as inflation and leptogenesis \cite{VanDong:2018yae}.
Note however that, as any non-supersymmetric electroweak gauge extension, 331 models do not -- \textit{per se} -- provide an explanation to the SM hierarchy problem. It is interesting to note, however, that even without imposing supersymmetry, these models can lead to new physics at the TeV scale, with potential new experimental signatures in several sectors, including the solution to the $g-2$ problem.

Due to the fact that $SU(2)_L$ gauge symmetry is promoted to $SU(3)_L$, the left-handed components of the fermion generations are arranged in the fundamental (or anti-fundamental) representation of $SU(3)_L$, i.e, triplets. Thus, the structure of the model is now based on triplets, instead of the usual doublets from the SM. A fundamental consequence of this extension is the prediction of the existence of other new gauge bosons, such as $Z^\prime$, $W^\prime$, as well as other exotics, such as doubly charged gauge bosons ($U^{++}$). Another change compared to the SM is that more than one scalar field is necessary to generate masses for all particles. Usually, 3-3-1 models employ three scalar triplets, with one of them being responsible for the Spontaneous Symmetry Breaking (SSB) of $SU(3)_L \times U(1)_X \to SU(2)_L \times U(1)_Y$  and the others driving electroweak symmetry breaking. After the $SU(3)_L \times U(1)_X$ symmetry is spontaneously broken due to the vacuum expectation value of one of the scalar triplets, namely $\chi$, a remnant $SU(2)_L \times U(1)_Y$ survives, recovering the SM gauge symmetry \cite{Borges_2016}. This change from $SU(2)_L$ doublets to $SU(3)_L$ triplets implies new fermions contributing to the gauge anomalies of the extended model. 
The electric charge operator that preserves the vacuum is given by, 

\begin{align*} 
\cfrac{Q}{e}&=\cfrac{1}{2} (\lambda_3+ \alpha \lambda_8)+ X \mathcal{I}\\
&= \operatorname{diag}\left(\cfrac{1}{2} \left(1+\beta\right) + X, \, \cfrac{1}{2} \left(-1+ \beta\right) + X, \,  - \beta + X\right),
\end{align*}
\noindent
where $\lambda_{3}=\operatorname{diag}\left(1,-1,0\right)$, $\lambda_{8}=\operatorname{diag}\left(1,1,-2\right) / \sqrt{3}$, and $\mathcal{I}$ are the diagonal generators of $\text{SU(3)}_\text{L}$ (Gell-Mann matrices), with the identity matrix acting as a generator of $\text{U(1)}_\text{X}$, respectively. X and $\alpha$ are free parameters \textit{a priori}, with different choices describing distinct models
under the 3-3-1 class, and $\beta =\sfrac{\alpha}{\sqrt{3}}$.

Since we need to recover the Standard Model spectrum, the first two components of the lepton triplet should have the same structure as the SM, represented by a neutrino and a charged lepton. We can fix $\beta=-(2X+1)$ with this choice. Thus, the third component of the triplet should have an electric charge quantum number equal to $3X+1$. The choice of X yields different 3-3-1 models, with the most famous cases being models with electrically charged ($X=0$) or neutral fields ($X = -1/3$). To cite a few of these models, we have, besides the original ones \cite{Singer:1980sw,Valle:1983dk}, the minimal 3-3-1 model \cite{Pisano:1991ee,Frampton:1992wt}, the 3-3-1 model with exotic charged leptons \cite{Ponce:2001jn,Ponce:2002fv,Anderson:2005ab,Cabarcas:2013jba}, the 3-3-1 model with right-handed neutrinos \cite{Hoang:1996gi,Hoang:1995vq}, the economical 3-3-1 model \cite{Dong:2006mg,Dong:2008ya,Berenstein:2008xg,Martinez:2014lta}, and 3-3-1 models with a left-handed neutral lepton \cite{Mizukoshi:2010ky, Catano:2012kw}. 

In the context of 3-3-1 models, the main contribution to the muon's $g-2$ comes from the gauge bosons of the model ($Z^\prime$ and $W^\prime$), where the neutral and charged currents are given by 
\begin{equation}
    {\cal L}^{NC} \supset
\bar{f}\, \gamma^{\mu} [g_{V}(f) + g_{A}(f)\gamma_5]\, f\,
Z'_{\mu},
\label{eqNC}
\end{equation}
\begin{equation}
{\cal L}^{CC} \supset - \frac{g}{\sqrt{2}}\left[
\overline{N_L}\, \gamma^\mu \bar{l_L}\, W^{\prime -}_\mu  \right],
\label{eqCC}
\end{equation}
 where $f$ is the leptonic triplet representation and $N_L$ represents the third component of the lepton triplet and 
\be
g_{V}(f) = \frac{g}{4 c_W} \frac{(1 -
4 s_W^2)}{\sqrt{3-4s_W^2}},\
g_{A}(f) = -\frac{g}{4 c_W \sqrt{3-4s_W^2}}, \nonumber
\label{gvga} 
\ee 
where $g$ is the weak coupling, $c_W = \cos{\theta_W}$, $s_W = \sin{\theta_W}$ are the cosine and sine of the Weinberg angle, respectively.

The masses of the $Z^\prime$ and $W^\prime$ gauge bosons are given by 
\begin{align}
M_{Z^{\prime}}^2 = \frac{g^2}{4(3-4s_W^2)}& \left(4 c_W^2 v_{\chi}^2 + \frac{v_{\rho}^2}{c_W^2}
+ \frac{v_{\eta}^2(1-2s_W^2)^2}{c_W^2} \right), \nonumber \\ 
M_{W^{\prime}}^2 &= \frac{g^2}{4}\left( v_{\eta}^2 + v_{\chi}^2 \right),
\label{Zprimemass}
\end{align} where $v_\rho$ and $v_\eta$ are the vacuum expectation values of the scalar triplets $\rho$ and $\eta$ driving electroweak symmetry breaking. The explicit dependence of the gauge boson masses on $v_\chi$ can be seen from \cref{Zprimemass}. This dependence will be important in what follows, as it will relate new gauge boson searches at high-energy colliders with the results obtained in our analysis.

Recently it has been shown that none of the above 3-3-1 models are capable of accommodating the $g-2$ anomaly, as a result of stringent collider bounds \cite{deJesus:2020ngn} (see \cite{Ky:2000ku,Kelso:2013zfa,Binh:2015cba,Cogollo:2017foz,DeConto:2016ith} for $\gmu$ studies in the context of 3-3-1 models). This result relies on the fact that the gauge boson masses are proportional to the energy scale that controls $SU(3)_L \times U(1)_X \to SU(2)_L \times U(1)_Y$ spontaneous symmetry breaking ($v_\chi$), which is the main parameter in the analysis. This dependence on the scale of SSB makes it possible to recast LHC lower bounds on the mass of the gauge bosons into constraints on the energy scale of 3-3-1 SSB. From LHC searches we have $v_\chi > 4 \, \mbox{TeV}$, which excludes most of the region where 3-3-1 models can account for the anomaly. 
Additionally, another recent study \cite{Cherchiglia:2022zfy} in the regime where 3-3-1 is an effective theory containing only scalars assessed the impact on $\gmu$, where they showed that, even including Barr-Zee type diagrams, it is not possible to explain $\gmu$, consistent with our statement.

It follows that, by themselves, the simplest 3-3-1 models cannot accommodate the muon magnetic moment anomaly, requiring the extension we consider here. The latter consists in embedding the simplified Lagrangians of Sec.\ref{secII} into 3-3-1 models by promoting the inert scalar doublet into a scalar triplet, to preserve the gauge invariance of the Lagrangian, while the other fields remain unchanged. The major change is that now the relevant parameter for the analysis will be the scale of symmetry breaking $v_\chi$, instead of the mass of the inert scalar, as in \cref{inertscalar,vectorlike}. The relationship between the inert scalar mass and $v_\chi$ involves scalar potential mixing terms of the type $(\phi^\dagger \phi)(\chi^\dagger \chi)$. After the SSB these generate a mass to $\phi$ proportional to $v_\chi$ \footnote{Mixing with other scalar fields could also generate mass terms for $\phi$. However, in 3-3-1 models it is usually assumed that the first SSB takes place at a higher energy scale than that which characterizes electroweak breaking.}. As a result the $(\phi^\dagger \phi)(\chi^\dagger \chi)$ term generates the most significant contribution to the inert scalar mass, which we will assume to be given as $m_\phi \sim \Lambda v_\chi$. 

\begin{figure}
    \centering
    \includegraphics[width=\columnwidth]{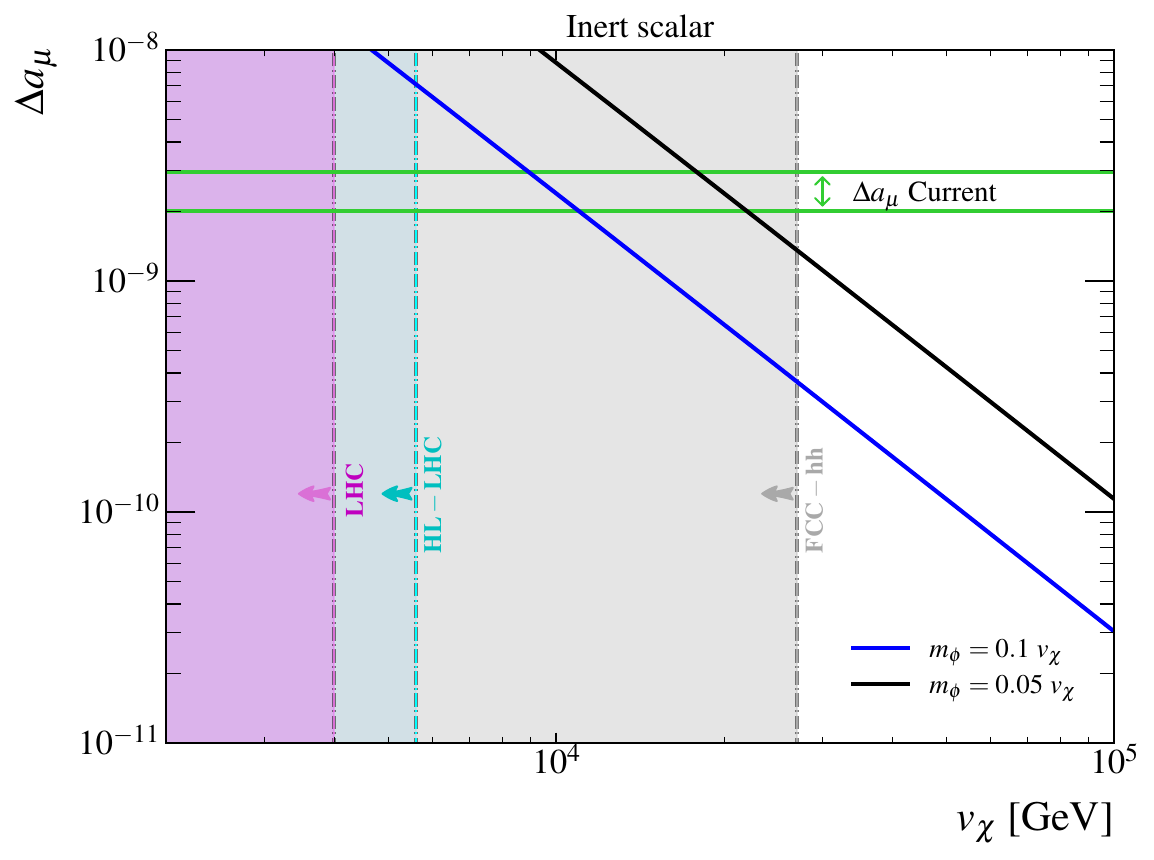}
    \caption{Graph illustrating the relationship between the anomalous magnetic moment ($\amu$) and the symmetry-breaking scale ($v_{\chi}$) for the $\gmu$ explanation. We assume $\lambda_{22} =4$ ($g^{s}_{22} =2$) and $g^{p}_{22} =0$ consistent with perturbative unitarity. The blue and black lines correspond to two distinct scenarios, characterized by scalar masses of $m_{\phi}= 0.05v_{\chi}$ and $m_{\phi}= 0.1v_{\chi}$ respectively. The vertical magenta, cyan, and gray regions give the current and projected exclusions from the LHC, HL-LHC, and FCC-hh, as indicated.}
    \label{figCla1}
\end{figure}

\begin{figure}
    \centering
    \includegraphics[width=1.\columnwidth]{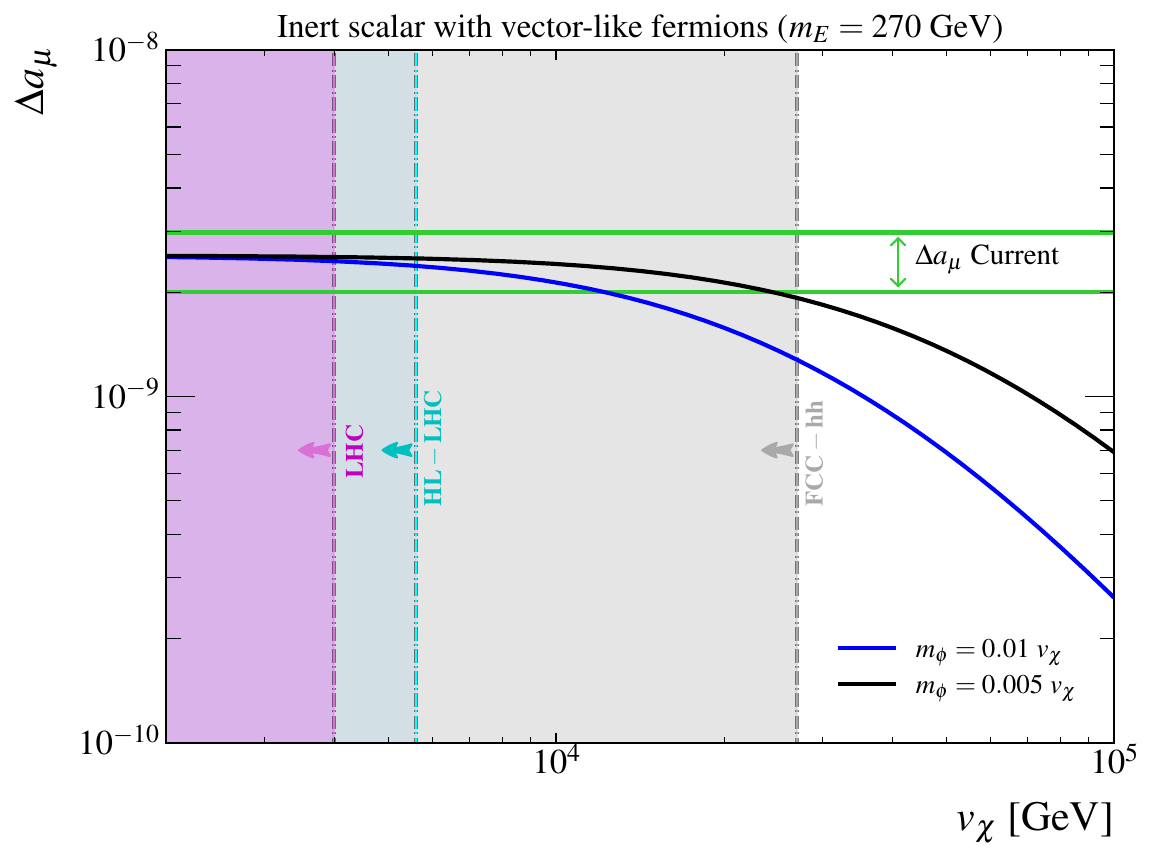}
    \caption{The plot illustrates the relationship between the anomalous magnetic moment of the muon ($\amu$) and the scale of symmetry breaking ($v_{\chi}$), providing an explanation for the $\gmu$ anomaly with an exotic charged lepton with a mass of $m_E=270$~GeV. We assume $\lambda_{12} =4$ ($g^{s(p)}_{12} =2$). The blue and black curves correspond to scalar masses of $m_{\phi}= 0.005v_{\chi}$ and $m_{\phi}= 0.01v_{\chi}$ respectively. The vertical regions are the current and projected exclusion areas resulting from high-energy collider searches at LHC (magenta), HL-LHC (cyan), and FCC-hh (gray).}    
    \label{figCla2}
\end{figure}

Our study of these simplified schemes embedded in 3-3-1 models yields basic features concerning the interactions involving the inert scalar. We have meticulously assessed their contribution to the muon's $\gmu$. Our results are summarized in \cref{figCla1,figCla2,mphixme}. The $5.0 \sigma$ deviation from the experimental value reported by Fermilab \cite{Muong-2:2023cdq} is represented by the region between the solid horizontal green lines in \cref{figCla1,figCla2} and also leads to the colored region in \cref{mphixme}. To get these results we have first scrutinized the scenario in which the inert scalar couples with the muon in the loop. Subsequently, we examined the case where it also couples with the vector-like fermion $E$. By varying the values for the masses of $\phi$ and E, we have explored which parameter values are most relevant for the $\gmu$ anomaly. 

Since our proposed scheme is embedded in a 3-3-1 setup, we can use the bounds from $Z^\prime$ searches in high-energy colliders to constrain $m_\phi$. In order to do this we take advantage of the mutual dependence on $v_\chi$, where the inert scalar mass is given by $m_\phi = \Lambda v_\chi$ while $m_{Z^\prime}$ is given by \cref{Zprimemass}. 

The high-energy bound from the LHC and the future expected sensitivities are represented by the vertical dot-dashed lines in \cref{figCla1,figCla2}, depicted in magenta, cyan, and gray color. These give the bounds from the LHC and the projections for the LHC upgrade to High-Luminosity (HL-LHC), as well as the Future Circular hadron Collider (FCC-hh), respectively. The bounds from the LHC were derived using an integrated luminosity of $\mathcal{L}=139 \, \mathrm{fb}^{-1}$ and a center-of-mass energy $\sqrt{s}=13$ TeV. In comparison, simulations for the HL-LHC utilized $\mathcal{L}=$ $3000 \, \mathrm{fb}^{-1}$ with $\sqrt{s}=14 \, \mathrm{TeV}$. The FCC-hh adopts the same luminosity level as the HL-LHC, but operates at a significantly higher center-of-mass energy of $\sqrt{s}=100 \, \mathrm{TeV}$, as referenced in \cite{Alves:2022hcp}. 
The limits are set at $4$,  $5.6$, and $27$ $\mathrm{TeV}$ for the LHC, HL-LHC, and FCC-hh, respectively. 

\cref{figCla1} represents the contribution to the $\Delta a_\mu$ coming from an inert scalar that couples with the muon as a function of the scale of 3-3-1 symmetry breaking. In this plot, we assumed the Yukawa coupling $\lambda_{22} =4$, so $g^{s}_{22} =2$ and $g^{p}_{22} =0$, consistent with perturbativity and unitarity. We took different configurations for the inert scalar mass and the 3-3-1 scale of symmetry breaking of the 3-3-1 symmetry, given by $m_{\phi} = 0.1 v_\chi$ and $m_{\phi} = 0.05 v_\chi$, represented by the blue and black solid lines, respectively. With this configuration, one can see that the region of $v_\chi$ and $m_\phi$, where one can accomodate the 5-sigma discrepancy claimed by Fermilab lies beyond the reach of current high-energy experiments. However, it could possibly be probed by future upgrades of the LHC or at high-energy lepton colliders such as ILC, CLIC, CEPC or FCC-ee.

On the other hand, \cref{figCla2} illustrates the contribution to the muon anomalous magnetic moment arising from an inert scalar which is also coupled with the exotic lepton E. Within this framework, we explore variations in the coupling constant of this interaction and the mass of E. This way we find the optimal parameter values that accommodate the current region of $\Delta a_{\mu}$. In \cref{figCla2} we display the predicted $\amu$ values versus $v_{\chi}$. One sees that $\gmu$ can indeed be understood from first principles with an exotic charged lepton with a mass of $m_E=270$~GeV. Our analysis adopts a specific value of $\lambda_{12} =4$ ($g^{s(p)}_{12} =2$), consistent with perturbative unitarity. The depicted blue and black curves give distinct scenarios corresponding to scalar masses of $m_{\phi}= 0.005v_{\chi}$ and $m_{\phi}= 0.01v_{\chi}$ respectively. This way we get further insights on how this scheme can reproduce the muon's anomalous magnetic moment. The result is similar to the case of the inert scalar, and one sees that current data cannot exclude the region of parameter space that can fit the anomaly. Future experiments, though, should be able to cover the parameter space that lies beyond the reach of current LHC searches, and produce a signal in future colliders, such as the FCC.

\section{\label{secIV} Conclusions}

This work investigated the possibility of inert scalars accounting for the recent $\gmu$ anomaly within simple schemes, as well as in the context of full-fledged 3-3-1 theories involving vector-like fermions. We considered the contributions of these particles to the muon anomalous magnetic moment and examined their viability in light of current experimental constraints. The simplified schemes with inert scalars and vector-like fermions could provide solutions to the $\gmu$ anomaly (\cref{inertscalar,vectorlike}). The associated regions of parameter space that could explain the anomaly are given in \cref{mphixme}. We found that vector-like fermion masses between $250 - 300$ GeV and inert scalar masses between $350 - 450$ GeV are sufficient to accommodate the anomaly.

Accounting for the $\gmu$ anomaly in the context of 3-3-1 gauge theories, however, is subject to limitations that follow mainly due to stringent collider bounds. These arise from experimental searches for new gauge bosons at high-energy colliders such as the LHC. These constraints apply to the new inert scalar triplet via the mass term, which is proportional to the 3-3-1 symmetry breaking scale ($v_\chi$) (see \cref{figCla1,figCla2}). Detailed analysis reveals that for the case of an Inert scalar without vector-like fermions one can accommodate the $\gmu$ anomaly while avoiding bounds from current data. In contrast, this is not possible for the case of the inert scalar plus vector-like fermions. Indeed, here the constraints from current collider data exclude a large area of parameter space which can accomodate the anomaly (magenta region). Projections taking into account the reach of LHC upgrades as well as those of new high-energy colliders, e.g., HL-LHC and FCC-hh, fully cover the parameter region compatible with the anomaly. This means that if this model is indeed the right answer to the anomaly, it should lead to a signal in the next generation of collider experiments.

\section*{Acknowledgements}

 FSQ This work was supported by Simons Foundation (Award Number:1023171-RC), FAPESP Grant 2018/25225-9, 
2021/01089-1, 2023/01197-4, ICTP-SAIFR FAPESP Grants 2021/14335-0, CNPq Grant 307130/2021-5, and ANID-Millennium Science Initiative Program ICN2019\textunderscore044
 YSV expresses gratitude to São Paulo Research Foundation (FAPESP) under Grant No. 2018/25225-9 and 2023/01197-4.
 JWFV is supported by Spanish grants PID2020-113775GB-I00 (AEI/10.13039/501100011033) and CIPROM/2021/054 (Generalitat Valenciana). 
 ASJ acknowledges support from Coordenaç\~ao de Aperfeiçoamento de Pessoal de N\'ivel Superior (CAPES) under Grant No. 88887.497142/2020-00.
 

\def\bibsection{\section*{References}}
\bibliography{bibliography}
\end{document}